\documentclass[aps,prl,twocolumn,showpacs,superscriptaddress,reprint]{revtex4-1}
\usepackage{graphicx}
\usepackage{natbib}
\usepackage{amsmath}
\usepackage{amssymb}
\usepackage{bm}
\usepackage{color}
\usepackage{lipsum}

\begin{document}

\title{Incomplete Protection of the Surface Weyl Cones of the Kondo Insulator SmB$_6$:\\ Spin Exciton Scattering.}

\author{G.A. Kapilevich}
\affiliation{Temple University, Philadelphia, Pa 19122, USA}
\author{P.S. Riseborough}
\affiliation{Temple University, Philadelphia, Pa 19122, USA}
\author{A.X. Gray}
\affiliation{Temple University, Philadelphia, Pa 19122, USA}
\author{M. Gulacsi}
\affiliation{Max Planck Institute for the Physics of Complex Systems, Dresden, Germany}
\author{Tomasz Durakiewicz}
\affiliation{Los Alamos National Laboratory, Los Alamos, NM 87545, USA}
\author{J.L. Smith}
\affiliation{Los Alamos National Laboratory, Los Alamos, NM 87545, USA}

\date{\today}

\begin{abstract}
The compound SmB$_6$ is a Kondo Insulator, where the lowest-energy bulk electronic excitations are spin excitons. It also has surface states that are subjected to strong spin-orbit coupling. It has been suggested that SmB$_6$ is also a topological insulator. Here we show that, despite the absence of time-reversal symmetry breaking and the presence of strong spin-orbit coupling, the chiral spin texture of the Weyl cone is not completely protected. In particular, we show that the spin-exciton mediated scattering produces features in the surface electronic spectrum at energies separated from the surface Fermi energy by the spin-exciton energy. Despite the features being far removed from the surface Fermi energy, they are extremely temperature dependent. The temperature variation occurs over a characteristic scale determined by the dispersion of the spin exciton. The structures may be observed by electron spectroscopy at low temperatures.

\end{abstract}

\maketitle


Heavy-fermion semiconductors, also known as Kondo insulators, are a family of semiconductors with extremely narrow gaps that are subjected to strong electron correlations \cite{RiseboroughA}. The compounds Ce$_3$Bi$_4$Pt$_3$, YbB$_{12}$ and SmB$_6$ can be considered as archetypal members of this class of materials. The Ce-based semiconductors have average occupancies of the atomic 4f electronic shell which are close to unity and, therefore, are close to the Kondo limit. The Ce 4f states are primarily linear superpositions of the non-magnetic 4f$^0$ and the moment carrying 4f$^1$ states, whereas the Yb-based compounds can be considered as the electron-hole symmetric partners of Ce involving the 4f$^{14}$ and 4f$^{13}$ configurations. The Sm-based compound SmB$_6$ is strongly mixed valent and involves the non-magnetic 4f$^6$ and the magnetic 4f$^5$ configurations that have small term splittings \cite{Martin,Nickerson}. The properties of the heavy-fermion semiconducting materials have been described by the Periodic Anderson Model in which the Fermi-level resides within the hybridization gap \cite{Riseborough1}. The strong correlations of Ce or Yb have been incorporated using the slave boson technique appropriate to configurations involving only one 4f electron or one 4f hole. Since the materials have strong correlations, like most heavy-fermions systems, they can be expected to be close to an instability to a magnetic phase. Therefore, these narrow-gap semiconductors should show magnetic fluctuations that are precursors of transitions to magnetic states \cite{Riseborough2,Riseborough3,Riseborough4}. This is analogous to the expectation that either paramagnons or antiparamagnons form in  paramagnetic metals as precursors to instabilities to ferromagnetic or antiferromagnetic states, respectively. The spin-exciton excitations were first predicted for Ce-based compounds, however, they were observed via inelastic neutron scattering experiments on SmB$_6$ by Alekseev {\it et al.} \cite{Alekseev} and on YbB$_{12}$ by Bouvet {\it et al.} \cite{Iga}. The relation of the spin-exciton excitations to quantum criticality has recently been strengthened by their observation in a heavy-fermion semiconductor CeFe$_2$Al$_{10}$ \cite{Mignot,Adroja}, which is intimately related to the antiferromagnetic system CeOs$_2$Al$_{10}$.
%

%
For SmB$_6$, the hybridization gap is of the order of 20 meV, whereas the spin-exciton dispersion relation is in the gap and has a minimum value of about 10 meV at the {\it R} point, ${2 \pi \over a}({1 \over 2},{1\over 2},{1\over 2})$, on the Brillouin zone boundary.  Recently, Fuhrman {\it et al.} \cite{Fuhrman} performed an extensive experimental investigation of the spin-excitons of SmB$_6$ and found that their dispersion relation has the same periodicity as the Brillouin zone, indicating that the spin-excitons are coherent. Furthermore, they found subsiduary minima in the dispersion relation at the {\it X} point ${2 \pi \over a}({1 \over 2},0,0)$. Fuhrman {\it et al.} found that the experimentally observed excitations are in excellent agreement with the spin-excitons of an Anderson Lattice Model that has been generalized to include direct 4f to 4f hopping processes.
\begin{figure}[!ht]
  \centering
  \includegraphics[width=6cm]{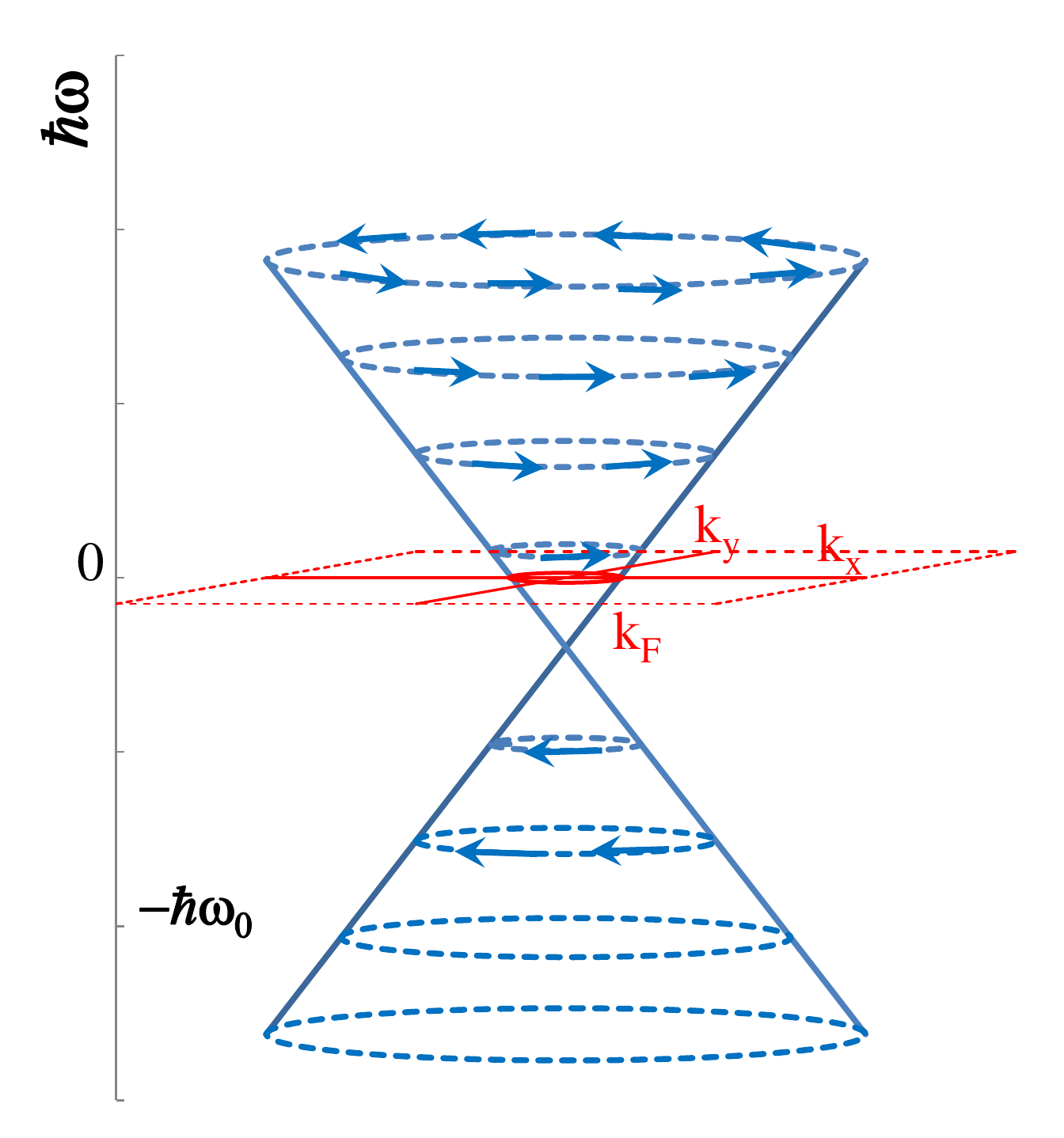}
  \caption{(color online) A sketch of a Weyl cone (blue) over the surface Brillouin zone (red) of SmB$_6$. The spin texture is denoted by the arrows. The red circle at $\omega=0$ depicts the surface Fermi-surface. }\label{SmB6NatCom}
\end{figure}
Since the resistivity increases with diminishing temperature but plateaus at low temperatures, it has been suggested that SmB$_6$ has surface states \cite{Kim,Zhang,Xu}. Like the 4f electrons of all lanthanide elements, the Sm 4f electrons are subject to strong spin-orbit coupling. The surface states are expected to experience a strong Rashba spin-orbit interaction that would lead to the direction of the spins locking to the momenta of the Weyl cones, as sketched in fig.(\ref{SmB6NatCom}). Such spin textures have been observed in non-correlated topological insulators, such as Bi$_2$Se$_3$ or Bi$_{1-x}$Sb$_{x}$ \cite{Xia,Hsieh,Hsieh2}, and also in SmB$_6$ \cite{Xu2}. It has been suggested that SmB$_6$ is a topological Kondo insulator \cite{Dzero,Takimoto}. Topological states are protected from scattering by non-magnetic impurities and may lead to dissipationless transport. However, magnetic impurities may result in spin-flip scattering, which can result in the reversal of momentum. Apart from the phonons, the spin-exciton excitations are expected to be the lowest energy bulk excitations of SmB$_6$. Here, we shall investigate the effect of the scattering of the surface states from the spin-flip scattering of bulk spin-excitons and examine their effect on the states of the Weyl cone.


The surface states are described by the Rashba Hamiltonian \cite{Rashba}
\begin{equation}
    H_{S-O} \ = \ c \ \hat{e}_z \ . \ \bigg( \ \underline{p} \ \wedge \ \underline{\sigma} \ \bigg)
\end{equation}
where $c$ will be the surface Fermi-velocity. The description can be thought of as originating from the Dirac equation in the massless limit, which reduces to two Weyl equations, each of which, separately, breaks inversion symmetry. The Rashba Hamiltonian can then be considered as the reduction of the Hamiltonian of the Weyl equation
\begin{equation}
    \bigg( \ i \ {\hbar \over c} \ {\partial \over \partial t} \ - \ \underline{\sigma} \ . \ \underline{p} \ \bigg) \ \psi \ = \ 0
\end{equation}
to two dimensions. This reduction is achieved via the substitution
\begin{equation}
    \underline{p} \ \rightarrow \ \hat{e}_z \ \wedge \ \underline{p} \ \ ,
\end{equation}
which eliminates $p_z$. The eigenstates of the Rashba Hamiltonian forms a Weyl cones which has a chiral spin texture. The eigenstates are found as
\begin{equation}
    \phi_{\tau,\underline{k}}(\underline{r}) = {1 \over \sqrt{2}} \ \left(
                                              \begin{array}{c}
                                                 - \tau \\
                                                {k_y  -  i  k_x \over \vert \underline{k} \vert} \\
                                              \end{array}
                                            \right)
      \ \exp \bigg[ - i ( \omega t -  \underline{k}  .  \underline{r}  )  \bigg] \ \ ,
\end{equation}
where the energy eigenvalues are $E_{\tau}(\underline{k}) \ = \ \tau \ \hbar \ c \ \vert \underline{k} \vert$ and where $\tau$ is the chiral index $\tau  = \pm 1$. The energy dispersion relations forms a Weyl cone with a vertex at the Weyl point. If the Fermi level is at an energy other than the Weyl point, one finds a circular surface Fermi-surface ring. Photoemission experiments reveal that the Weyl cones in SmB$_6$ are centered on the {\it X} and the $\Gamma$ points of the surface Brillouin zone and that the surface Fermi-surface rings, respectively, have radii of 0.29 and 0.09 {\AA}$^{-1}$ \cite{Neupane,Jiang,Xu}. Tunneling measurements have tentatively identified a spectroscopic feature at -0.5 meV below the Fermi energy as the Weyl point \cite{Park}. Photoemission experiments with circularly polarized light \cite{Xu2} show circular dichroism (left minus right circularly polarized light) in which there is an asymmetry in intensity between the lobes at the +{\it X} and -{\it X} points. This is a consequence of the chiral spin textures.
\begin{figure}[!ht]
  \centering
  \includegraphics[width=6cm]{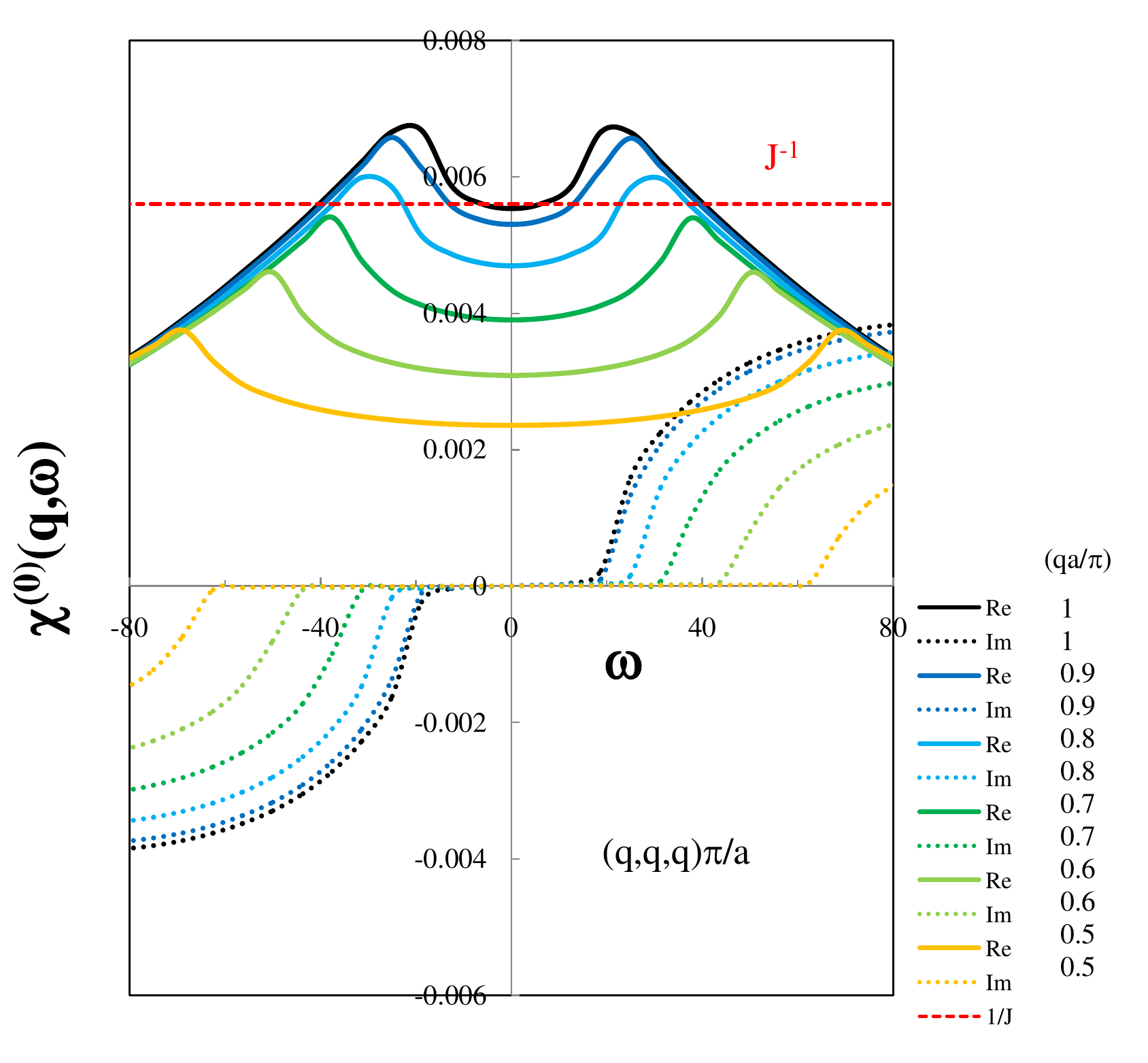}
  \caption{(color online) The real (solid) and imaginary (dashed) parts of the dynamic quasi-particle susceptibility $\chi^{(0)}(\underline{q},\omega)$ of the Anderson Lattice Model with the chemical potential $\mu$ in the gap,  for various wave vectors $\underline{q}$. The wave vector $\underline{q}$ is along the body diagonal and the $q$ values are shown in the legend. }\label{semichiq}
\end{figure}
The bulk spin-exciton excitations couple to the surface states via a Heisenberg exchange interaction $J'$,
which, when expressed in terms of the Rashba states, takes the form
\begin{widetext}
\begin{eqnarray}
    \hat{H}_{int} & = & + \ {J' \over 4 \ N} \ \sum_{\underline{q},\underline{k}_{\|},\tau,\tau'} \ \bigg[ \ \tau' \ \exp [ -i\varphi_{\underline{k}+\underline{q}} ] \ S^{+}_{-\underline{q}} \ + \ \tau \ S^{-}_{-\underline{q}} \ \exp[ + i \varphi_{\underline{k}} ] \nonumber \\
     && \ \ \ \ \ \ \ \ \ - \ \bigg( \ \tau \ \tau' \ - \ \exp [ -i(\varphi_{\underline{k}+\underline{q}}-\varphi_{\underline{k}}) ] \ \bigg) \ S^z_{-\underline{q}} \ \bigg] \ c^{\dag}_{\underline{k}+\underline{q},\tau} \ c_{\underline{k},\tau'}
\end{eqnarray}
\end{widetext}
The interaction describes includes a coupling between the states of the upper and lower part of the Weyl cone. The phase $\varphi_{\underline{k}}$ term depends on the orientation of the spin for state $\underline{k}$ and is given by
\begin{equation}
    \exp [ i \varphi_{\underline{k}} ] \ = \ \bigg( {k_y + i k_x \over \vert \underline{k} \vert} \bigg) \ \ .
\end{equation}

Because the system is both cubic and paramagnetic, the bulk magnetic susceptibility is isotropic in spin space and is given by the expression
\begin{equation}
    \chi^{\alpha,\alpha}(\underline{q},\omega) \ = \ { \chi^{(0)}(\underline{q},\omega) \over 1 \ - \ J \ \chi^{(0)}(\underline{q},\omega)} \ \ ,
\end{equation}
where $J$ is the effective antiferromagnetic exchange interaction, and $\chi^{(0)}(\underline{q},\omega)$ is the f electron quasi-particle susceptibility, shown in fig.(\ref{semichiq}) for a hybridization gap of $\Delta =20$ meV.  Using a value of the exchange interaction of $J \approx 200$ meV, the static uniform susceptibility is estimated as $3.6 \times  10^{-4}$ emu/mole, but the unenhanced static staggered susceptibility rises to $2.8 \times \ 10^{-3}$ emu/mole at the {\it R} point, indicating a strong tendency for Neel ordering.  Since the spin-exciton dispersion relation is relatively flat, we believe that the non-interacting susceptibility needs to be less $q$-dependent if it is to describe SmB$_6$.  Due to time reversal invariance, the real part of $\chi^{(0)}(\underline{q},\omega)$ is an even function of $\omega$, and the imaginary part is an odd function. Also note that the real part exhibits a pronounced minimum at $\omega=0$. The imaginary part of the susceptibility is zero within the gap. The spin-exciton energy $\omega_{\underline{q}}$ is given by the solution of
\begin{equation}
    1 \ = \ J \ \chi^{(0)}(\underline{q},\omega_{\underline{q}})
\end{equation}
which, since the imaginary part of $\chi^{(0)}(\underline{q},\omega_{\underline{q}})$ is required to vanish, must lie outside the electron-hole gap. That is, the spin-exciton occurs within the semiconducting gap. For the parameters chosen in the figure, one finds that for $q$ values smaller than 0.7${\pi \over a}$ the spin-exciton has merged with the electron-hole continuum and exists as a broadened resonance or anti-paramagnon.  Furthermore, one sees that if $\omega_{\underline{q}}$ approaches zero at a wavevector $\underline{Q}$, the criterion reduces to
\begin{equation}
    1 \ = \ J \ \chi^{(0)}(\underline{Q},0) \ \ .
\end{equation}
The resulting divergence of the static susceptibility $\chi^{\alpha,\alpha}(\underline{Q},0)$ signals the instability of the system to a static spin-density wave with wave vector $\underline{Q}$. Hence, the spin-exciton in SmB$_6$ can be regarded as a precursor excitation for an antiferromagnetic instability that occurs at a nearby point in parameter phase space.
\begin{figure}[!ht]
  \centering
  \includegraphics[width=6cm]{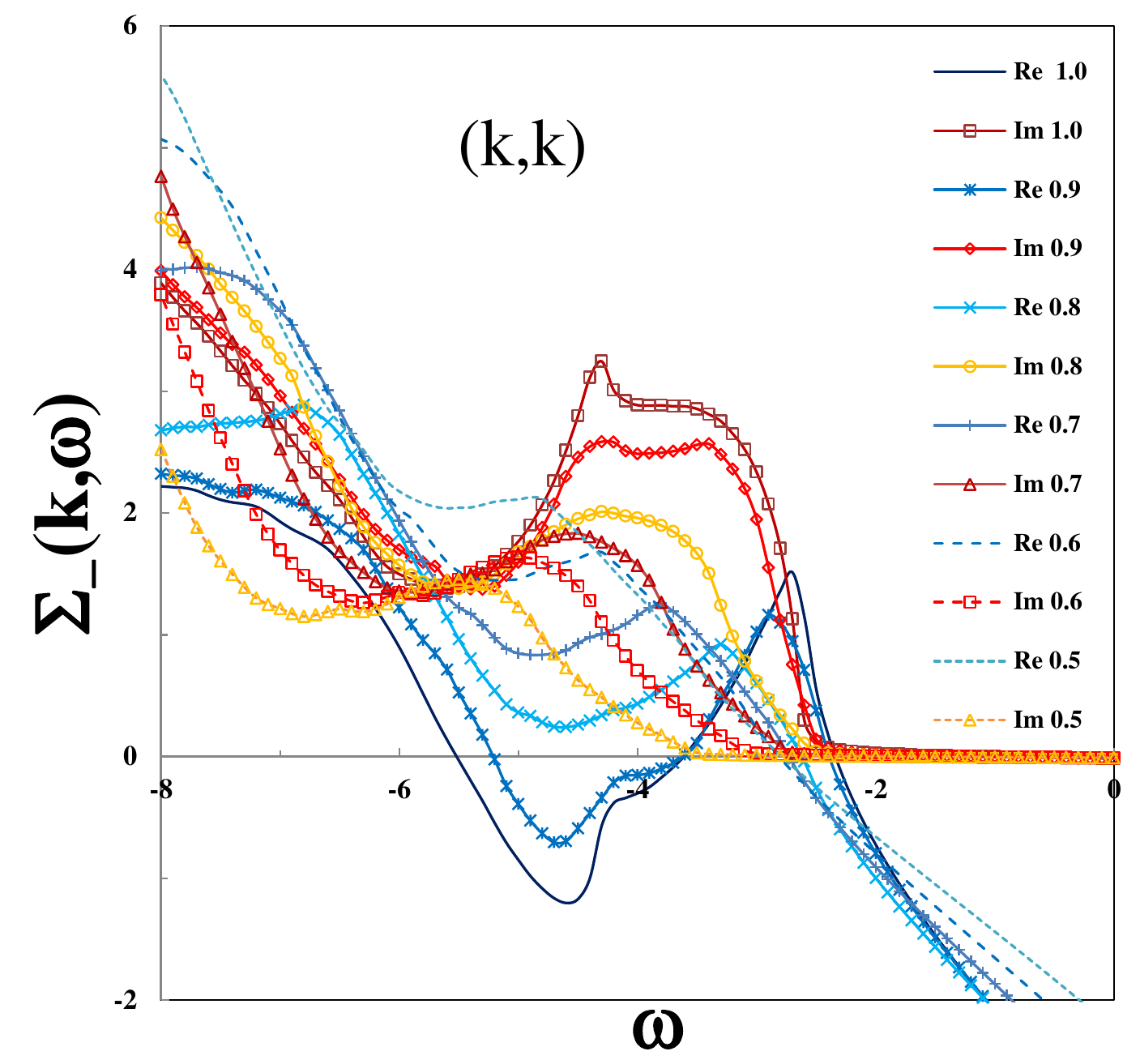}
  \caption{(color online) The frequency dependence of the real and imaginary parts of the self energy for a system close to a bulk Quantum Critical Point, for various wave vectors. The wave vectors $\underline{k}$ are along the diagonal of the surface Brillouin zone and the values of the components are given in the legend in units of ${\pi\over a}$. }\label{selfQC}
\end{figure}
The spectrum of magnetic excitations is given by the imaginary part of the susceptibility. Within the gap, the imaginary part of the susceptibility $\chi^{\alpha,\alpha}(\underline{q},\omega)$ has a delta function contribution from the spin-exciton, which is given by
\begin{equation}
     \Im m \ \chi^{\alpha,\alpha}(\underline{q},\omega) \ = \ {1 \over J} \ sign(\omega) \ \pi \ \delta\bigg( \ 1 \ - \ J \ \chi^{(0)}(\underline{q},\omega) \ \bigg)
\end{equation}
and reduces to
\begin{equation}\label{11}
    \Im m \ \chi^{\alpha,\alpha}(\underline{q},\omega) \ = \ {sign(\omega)\over Z_{\underline{q}}} \ \pi \ \delta( \ \omega \ - \ \omega_{\underline{q}} \ ) \ \ .
\end{equation}
For SmB$_6$, the spin-exciton excitation energy $\omega_{\underline{q}}$ has been found to disperse from 14 meV to about 10 meV, and has its minimum excitation energy at the {\it R} point. The dimensionless factor $Z_{\underline{q}}$ is defined as a derivative with respect to $\omega$
\begin{equation}
   Z_{\underline{q}} \ = \ \bigg{\vert} \ J^2 \ \bigg({\partial \over \partial \omega} \ \chi^{(0)}(\underline{q}, \omega)\bigg)\bigg{\vert}_{\omega=\omega_{\underline{q}}} \ \bigg{\vert}
\end{equation}
evaluated at $\omega_{\underline{q}}$. The factor of $Z_{\underline{q}}$ is responsible for the peaking of the intensity of the spin-exciton for momentum transfers $\underline{q}$ near the points $({1 \over 2},{1 \over 2},{1 \over 2})$ and $({1\over 2},0,0)$. In particular, for low values of $\omega_{\underline{q}}$ it varies linearly with $\omega_{\underline{q}}$  and rapidly grows as $\omega_{\underline{q}}$ approaches the threshold of the electron-hole continuum. Thus, as the spin-exciton mode softens when quantum criticality is approached, its intensity grows.


An electron in a surface state with momentum $\underline{k}$ has a self energy due to the emission and absorption of the spin-exciton excitations, which is determined to be
\begin{widetext}
\begin{eqnarray}
    \Sigma_{\tau}(\underline{k},\omega) & = & {J'^2\over 24 \ N} \ \sum_{\underline{q},\tau'} \ \bigg[  3  -  \tau  \tau' \cos\bigg(\varphi_{\underline{k}-\underline{q}}-\varphi_{\underline{k}}\bigg) \bigg] \ \int_0^{\infty} d\omega' \ \bigg({\Im m \ \chi^{\alpha,\alpha}(\underline{q},\omega') \over \pi}\bigg) \nonumber \\
    && \times \bigg[ \ { 1 - f_{\tau',\underline{k}-\underline{q}_{\|}} + N(\omega') \over \omega +  \mu - E_{\tau'}(\underline{k}-\underline{q}_{\|}) - \omega' + i \eta} +  { \ f_{\tau',\underline{k}-\underline{q}_{\|}} + N(\omega') \over \omega + \mu - E_{\tau'}(\underline{k}-\underline{q}_{\|}) + \omega' - i \eta} \ \bigg] \ \ ,
\end{eqnarray}
\end{widetext}
where we have assumed that the component of $\underline{q}$ perpendicular to the surface is not conserved. The above expression is related to the self energy due to spin fluctuations in paramagnetic materials close to quantum critical points. The integral over $\omega'$ can be simply evaluated by using eqn.(\ref{11}). The electronic spectrum of the surface states, $A_{\tau}(\underline{k},\omega)$, can be calculated from $\Sigma_{\tau}(\underline{k},\omega)$ (see Additional Material).

We have evaluated the surface one-electron self energy for a system in which the bulk is close to a quantum critical point. The results are shown in fig.(\ref{selfQC}) for a spin exciton with a minimum excitation energy of $\omega_0  = 2.5$, $\mu=2.0$, $(J'/J)^2={ \pi \over 600}$, and for wave vectors along the diagonal of the surface Brillouin zone. It is seen that the real part of the self energy has a kink near $\omega \ = \ - \ \omega_0$ at which point the imaginary part rapidly increases with decreasing frequency. The self energy has a slight $\underline{k}$ dependence, reflecting a nearly nesting condition, and results in hot patches in the near Fermi energy portions of the Weyl cone.

For SmB$_6$, the spin exciton dispersion relation has a maximum of a few meV. The imaginary part of the self energy is maximized when the spin exciton has a minimal dispersion and is completely within the gap, since this maximizes the intensity of the magnetic excitations and increases the phase space for scattering. An analytic expression for the self energy in the limit of dispersionless spin-excitons is given in the supplementary material. As seen in fig.(\ref{anayl}), the resulting quasi-particle scattering rate jumps abruptly at the excitation energy $\omega = -  \omega_0$. The magnitude of the jump is determined by the surface density of states at the Fermi energy. The abruptness of the jump is due to our neglect of the dispersion $\omega_d$ in the spin exciton spectrum, and, as seen in fig.(\ref{selfQC}), for finite $\omega_d$  the rapid increase will take place over an energy range given by by $\omega_d$. The real part of the self energy shows a sharp cusp in the vicinity of $- \omega_0$. The resulting surface quasi-particle density of states is shown in fig.(\ref{specanayl}), which shows a V-like variation characteristic of the Weyl cone where the density of states goes to zero at the energy of the vertex. However, at the spin-exciton energy, the V-like variation abruptly ceases. The abrupt drop in the spectral density is primarily due to the divergence in the quasi-particle wave function renormalization
that occurs at the cusp and is caused by resonant scattering with the spin excitons.
\begin{figure}[!ht]
  \centering
  \includegraphics[width=6cm]{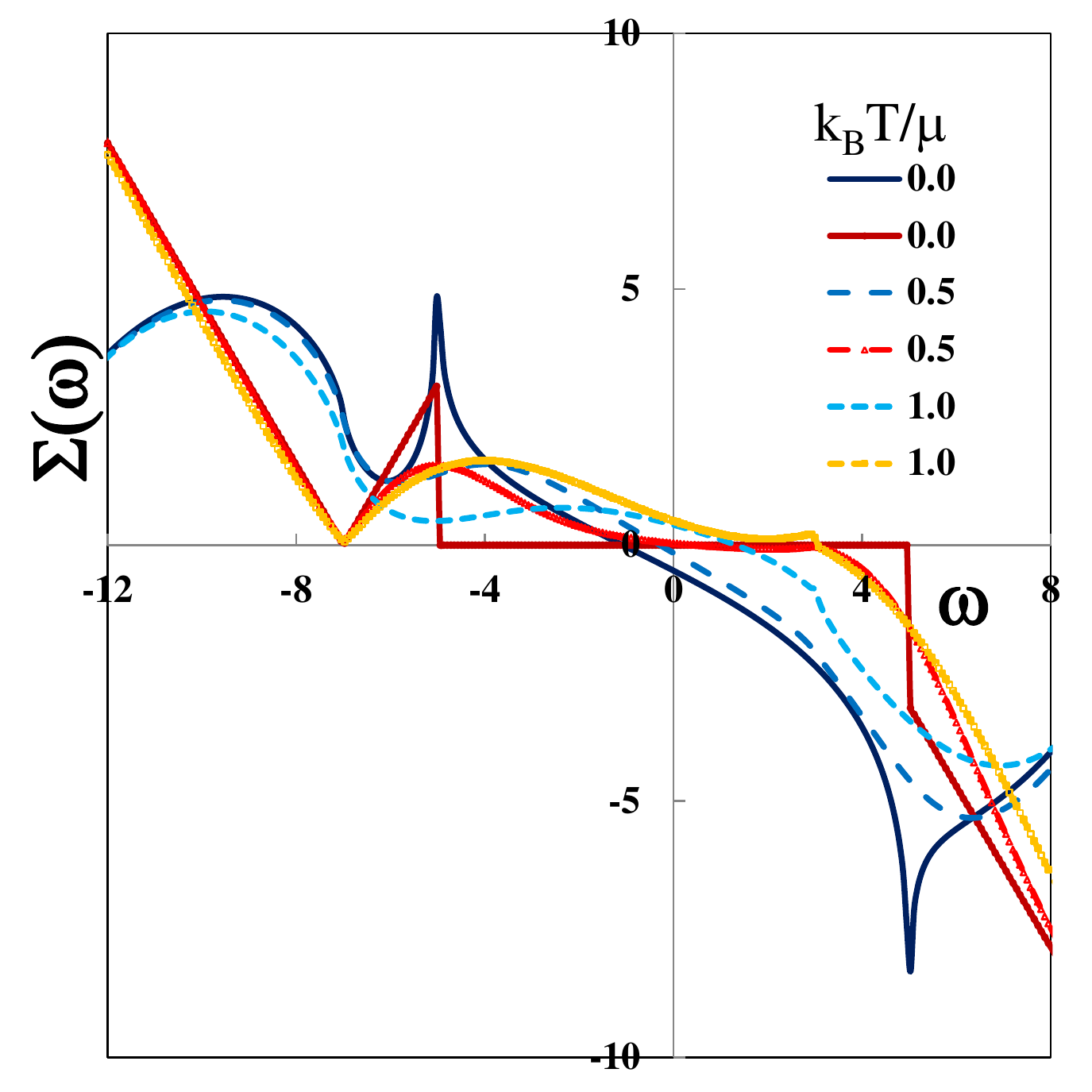}
  \caption{(color online) The real and imaginary parts of the self energy at various temperatures (in units of meV) calculated for a flat spin-exciton dispersion relation of $\omega_0=5$ meV, and the chemical potential is given by $\mu=2$ meV. The imaginary parts are denoted by the red lines. The imaginary part of the self energy is zero in the frequency between $+\omega_0$ and $-\omega_0$. The large features in the real part of the self energy at the edges of the frequency range are due to the proximity of the band edges of the cone states. The cusps in the real part of the self energy at $\omega = \pm \omega_0$ are seen to rapidly wash out as $T$ increases.}\label{anayl}
\end{figure}
\begin{figure}[!ht]
  \centering
  \includegraphics[width=6cm]{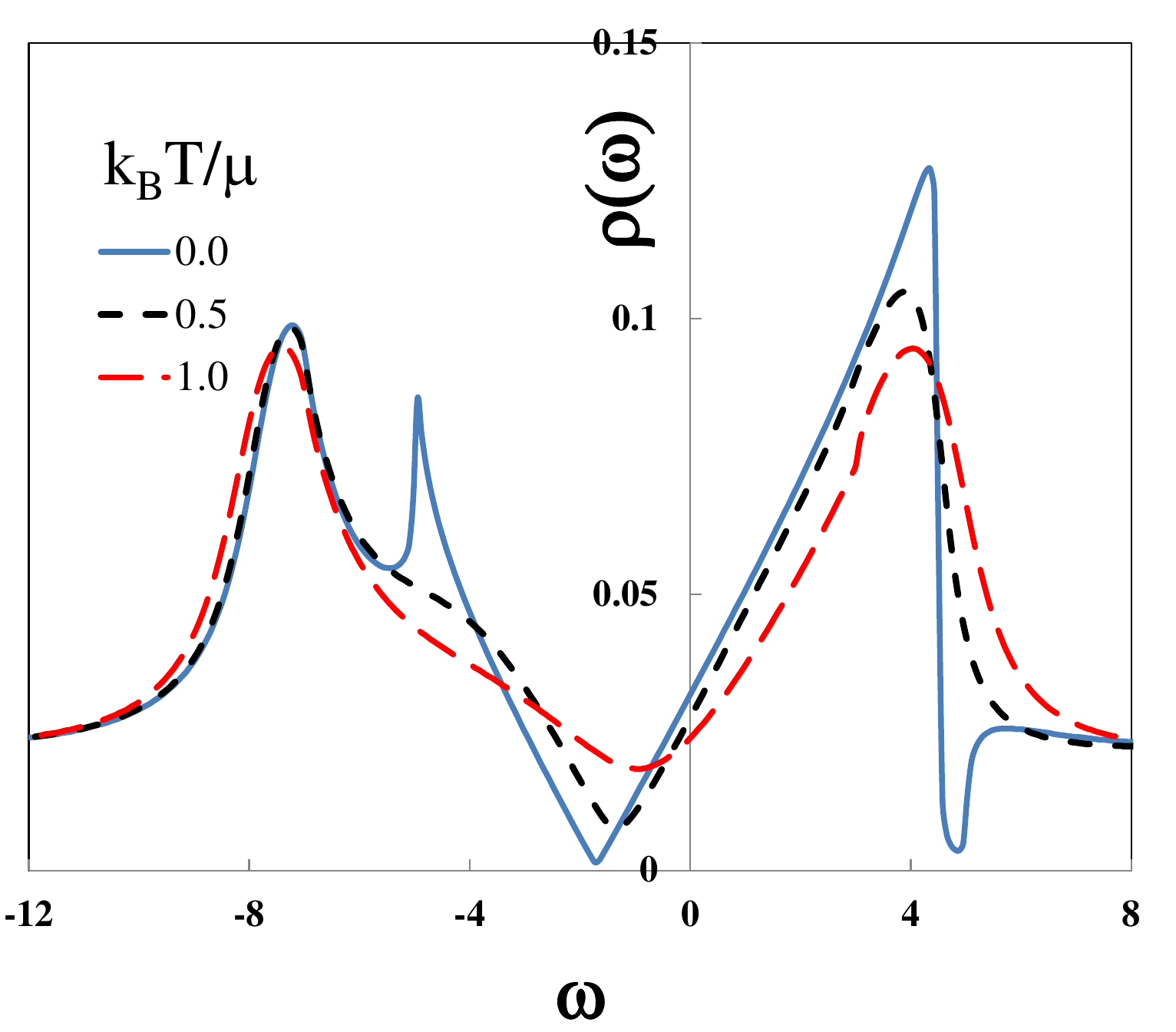}
  \caption{(Color on line) The calculated angle integrated surface photoemission spectra at various temperatures. The spectrum was calculated for the same parameters as used in fig.(\ref{anayl}). The density of states is calculated by assuming that, in addition to the self energy from the emission and absorption of spin-exciton excitations, there is an extremely small concentration of magnetic impurities. It is seen that distinct features near $\omega=\pm \omega_0$ rapidly disappear as $T$ is increased.}\label{specanayl}
\end{figure}
The temperature variation of this feature is not characterized by $\hbar \omega_0$ but is due to the surface electrons in the vicinity of the Fermi energy. The temperature scale is actually set by the dispersion $\omega_d$ of the spin-exciton mode.


In conclusion, Weyl cones in a strongly correlated Kondo insulator, in which there are low-energy spin exciton excitations, are neither protected by topology nor by the spin-orbit coupling despite the absence of broken time-reversal symmetry. We have found that at zero temperature, the imaginary part of the self energy exhibits a non-analytic behavior at the spin-exciton energy. The non-analytic behavior is a reflection of the non-analytic behavior of the unrenormalized density of states of the Weyl cone. The non-analytic behavior of the imaginary part of the self energy leads to an anomaly in the dispersion relation of the surface quasi-particles states that is responsible for structure in the electronic spectrum. Although this feature is at $\omega_0$ and is far removed from the surface Fermi energy, it is extremely temperature dependent and washes out rapidly as the temperature is increased. The rapid temperature dependence originates from a virtual process that involves states right at the surface Fermi energy, and the temperature scale is set by the dispersion in the spin-exciton energy. The spin-exciton induced structure in the surface electronic spectrum may be measured by high-resolution ARPES measurements or by tunneling experiments at low temperatures.

{\bf{Acknowledgements.}} The work at Temple was supported by an award from the US Department of Energy, Office of Basic Energy Sciences, via the award DE-FG02-01ER45872. PSR would like to acknowledge stimulating conversations with Pedro Schlottman, Collin Broholm, Wes Fuhrman, Laura Greene and Wan-Kyu Park. TD acknowledges the NSF IR/D program and the Department of Energy, Office of Basic Energy Sciences, Division of Material Sciences, Los Alamos National Laboratory.\\

\section{Supplementary Material}

\subsection{The Spin Exciton Weyl State Coupling}

The transformation between the decoupled spin and momentum eigenstates and the Rashba energy eigenstates is given by
\begin{eqnarray}
    \phi_{\uparrow,\underline{k}}(\underline{r}) & = &  {1 \over \sqrt{2}} \ \bigg( \ \phi_{-,\underline{k}}(\underline{r}) \ - \ \phi_{+,\underline{k}}(\underline{r}) \ \bigg) \nonumber \\
    \phi_{\downarrow,\underline{k}}(\underline{r}) & = &  {1 \over \sqrt{2}} \ \bigg( \ \phi_{-,\underline{k}}(\underline{r}) \ + \ \phi_{+,\underline{k}}(\underline{r}) \ \bigg) \ \exp [ i  \varphi_{\underline{k}} ]
\end{eqnarray}
where the phase factor is given by
\begin{equation}
    \exp [ i \varphi_{\underline{k}} ] \ = \ \bigg( {k_y + i k_x \over \vert \underline{k} \vert} \bigg) \ \ .
\end{equation}
The phase $\varphi_{\underline{k}}$ is simply related to the spin orientation of an electron with momentum $\underline{k}$.

The bulk spin-exciton excitations are assumed to couple to the surface states via a Heisenberg exchange interaction
\begin{equation}
    \hat{H}_{int} \ = \ - \ {J' \over N} \ \sum_{\underline{q},\underline{k}_{\|},\sigma,\sigma'} \ \underline{S}_{-\underline{q}} \ . \ \underline{s}_{\sigma,\sigma'} \ c^{\dag}_{\underline{k}+\underline{q},\sigma'} \ c_{\underline{k},\sigma}
\end{equation}
where $\underline{s}$ is the vector spin operator for the surface electronic states and $\underline{S}$ is the spin-operator which describes the spin excitations of the bulk material. When expressed in terms of the Rashba states, the interaction takes the form
\begin{widetext}
\begin{eqnarray}
    \hat{H}_{int} & = & + \ {J' \over 4 \ N} \ \sum_{\underline{q},\underline{k}_{\|},\tau,\tau'} \ \bigg[ \ \tau' \ \exp [ -i\varphi_{\underline{k}+\underline{q}} ] \ S^{+}_{-\underline{q}} \ + \ \tau \ S^{-}_{-\underline{q}} \ \exp[ + i \varphi_{\underline{k}} ] \nonumber \\
     && \ \ \ \ \ \ \ \ \ - \ \bigg( \ \tau \ \tau' \ - \ \exp [ -i(\varphi_{\underline{k}+\underline{q}}-\varphi_{\underline{k}}) ] \ \bigg) \ S^z_{-\underline{q}} \ \bigg] \ c^{\dag}_{\underline{k}+\underline{q},\tau} \ c_{\underline{k},\tau'} \ \ .
\end{eqnarray}
\end{widetext}
The interaction describes includes a coupling between the states of the upper and lower parts of the Weyl cone. The coherence factor in front of the $S^{z}_{-\underline{q}}$ term depends on the relative orientation of the initial and final spin state.

\subsection{The Dispersionless Limit}

For SmB$_6$, the spin exciton dispersion relation has a maximum of a few meV. The imaginary part of the self energy maximized when the spin-exciton has a minimal dispersion and is completely within the gap because this maximizes the intensity of the magnetic excitations and increases the phase space for scattering.
The integral over $\omega'$ in the expression for the self energy can be simply evaluated by using eqn.(11). This results in the expression
\begin{widetext}
\begin{eqnarray}
    \Sigma_{\tau}(\underline{k},\omega) & = & {J'^2\over 8 \ N} \ \sum_{\underline{q},\tau'} \ {1\over Z_{\underline{q}}} \ \bigg[  3 - \tau  \tau' \cos \bigg(\varphi_{\underline{k}-\underline{q}}-\varphi_{\underline{k}}\bigg)  \bigg] \nonumber \\
    && \times \bigg[ \ { 1 - f_{\tau',\underline{k}-\underline{q}_{\|}} + N(\omega_{\underline{q}}) \over \omega + \mu - E_{\tau'}( \underline{k}-\underline{q}_{\|}) - \omega_{\underline{q}} + i \eta} \ + {  f_{\tau',\underline{k}-\underline{q}_{\|}} + N(\omega_{\underline{q}}) \over \omega + \mu - E_{\tau'}(\underline{k}-\underline{q}_{\|}) + \omega_{\underline{q}} - i \eta} \ \bigg] \ \ .
\end{eqnarray}
\end{widetext}
Hence, the self energy has an explicit dependence on the square of the ratios of the surface to bulk exchange interactions. One can find an approximate analytic expression for the self energy of the surface states in the limit of a dispersionless spin-exciton by using a continuum model for the density of states per surface atom for the Weyl cone. The density of states $\rho_0(\epsilon)$, is given by
\begin{equation}
\rho_0(\epsilon) \ = \ {\epsilon \ a^2 \over 2 \ \pi \ c^2} \ \bigg[ \ 2 \ \Theta(\epsilon) \ - \ \Theta(\epsilon-{\Delta \over 2}) \ - \ \Theta(\epsilon + {\Delta \over 2}) \ \bigg] \ \ .
\end{equation}
The value of $c$, the surface Fermi-velocity, is chosen such that the surface dispersion relation joins the bands at the surface Brillouin zone boundary, so $c/a={\Delta \over 4 \ \sqrt{2}}$. One finds that the real and imaginary parts of the self energies are independent of $\underline{k}$ and, taking $x_{\pm}=\omega+\mu\pm\omega_0$, can be evaluated at $T=0$ as
\begin{widetext}
\begin{eqnarray}
\Re e \ \Sigma_{\tau}(\omega) & = & \bigg({3 \ J'^2 \ a^2 \over 16  \ \pi \ c^2 \ Z}\bigg) \ \bigg[   x_- \ln \bigg{\vert}  { \omega -  \omega_0 \over x_-  -  {\Delta \over 2} }  \bigg{\vert} -  x_+  \ln \bigg{\vert}  { \omega  +  \omega_0 \over x_+ } \bigg{\vert}  \nonumber
 +  x_+ \ln \bigg{\vert} {x_+ \over x_+  + {\Delta \over 2} } \bigg{\vert} \ \ \bigg]
\end{eqnarray}
and
\begin{eqnarray}
\Im m \ \Sigma_{\tau}(\omega) & = &\bigg({3  \pi  J'^2 a^2 \over 16  \pi  c^2 \ Z}\bigg) \ \bigg[ \ -  x_- \bigg( \ \Theta({\Delta \over 2}-x_- ) \ - \ \Theta(\omega_0-\omega) \ \bigg)  + x_+ \bigg( \ \Theta (x_+) \ - \ \Theta(\omega+\omega_0) \ \bigg) \nonumber \\
& & \ \ \ \ \ \ \ \ \ \ \ \ \ \ \ \ \ \ \ \ \ \ - x_+ \bigg( \ \Theta({\Delta \over 2}+x_+) \ - \ \Theta( x_+) \ \bigg) \ \bigg] \ \ .
\end{eqnarray}
\end{widetext}
The overall dimensionless multiplicative factor in the expression for the self energy has an order of magnitude that is given by
\begin{equation}
 \bigg({3  J'^2 a^2 \over 16  \pi  c^2 Z}\bigg) \ = \ {300 \over 4 \ \pi} \ \bigg({J' \over J}\bigg)^2 \ {\Delta \over \omega_0}
\end{equation}
The ratio ${J'\over J}$, the ratio of the Kondo interaction $J' \sim \Delta$ to the strength of the RKKY interaction [1] $J$ required to produce the magnetic exciton in SmB$_6$  $J \sim 200$, is estimated to be one tenth. This estimate seems reasonable since SmB$_6$ is mixed valent and not close to the Kondo limit [2].
The electronic spectrum of the surface states is given by
\begin{widetext}
\begin{equation}
    A_{\tau}(\underline{k},\omega) = \pm {1\over \pi} \ { \Im m \ \Sigma_{\tau}(\underline{k},\omega) \over [ \omega + \mu -  E_{\tau}( \underline{k}) - \Re e \ \Sigma_{\tau}(\underline{k},\omega) \ ]^2 + [ \Im m \ \Sigma_{\tau}(\underline{k},\omega)  ]^2}
\end{equation}
\end{widetext}

\subsection{References}

[1] S. Doniach, Physica B \& C, {\bf 91}, 231-234 (1977).\\

[2] S. Doniach, Phys. Rev. B, {\bf 35}, 1814-1821 (1987).\\

\end{document}